\documentclass{aa}
\usepackage{psfig}
\usepackage{natbib}
\begin{document}
\title{Energy release due to antineutrino untrapping and
diquark condensation in hot quark star evolution}
\titlerunning{Energy realease due to antineutrino untrapping...}
%
   \author{D.N. Aguilera \inst{1,2}
        \and D. Blaschke \inst{1,3}
        \and H. Grigorian \inst{1,4}
\thanks{Research supported by DFG under grant no. 436 ARM 17/1/00}
}
\offprints{D. Blaschke}

   \institute{Fachbereich Physik, Universit\"at Rostock,
        Universit\"atsplatz 1, D--18051 Rostock, Germany\\
        \and Instituto de F\'{\i}sica Rosario, Bv. 27 de febrero 210 bis,
        2000 Rosario, Argentina\\
         email: deborah@darss.mpg.uni-rostock.de\\
        \and Bogoliubov Laboratory for Theoretical Physics, JINR Dubna,
        141980 Dubna, Russia\\
        email: david@thsun1.jinr.ru
        \and Department of Physics, Yerevan State University, Alex
        Manoogian Str. 1, 375025 Yerevan, Armenia\\
        email: hovik@darss.mpg.uni-rostock.de
             }


\abstract{
We study the consequences  of antineutrino trapping in hot quark matter
for quark star configurations with possible diquark condensation.
Due to the conditions of charge neutrality and $\beta$-equilibrium
the flavor asymmetry increases with the number density of trapped
antineutrinos. Above a critical value of the antineutrino chemical
potential of $30$ MeV diquark condensation is inhibited at low densities
and a two-phase structure emerges:
a superconducting quark matter core surrounded by a shell of normal quark
matter.
When the quark star cools down below a temperature $T \sim 1$ MeV, the
mean free path of antineutrinos becomes larger than the thickness of the
normal quark matter shell so that they
get untrapped within a sudden process.
By comparing the masses of configurations with the same baryon number we
estimate that the release of energy due to the antineutrino untrapping
transition can be in the range of $10^{51} \div 10^{52}$ erg.
}

\maketitle


\section{Introduction}
\label{sec:intro}

The engine which drives supernova explosions and gamma ray bursts being
among the most energetic phenomena in the universe remains still puzzling
[\cite{Piran:2001da}].
The phase transition to a quark matter phase may be
a mechanism that could release such an amount of energy
[\cite{Drago:1997tn}, \cite{Berezhiani:2002ea}].
It has been proposed that due to the Cooper instability in dense Fermi gases
cold dense quark matter shall be in the color superconducting state with
a nonvanishing diquark condensate [\cite{Alford:2000sx},
 \cite{Blaschke:2001uj}].
The consequences of diquark condensation for the cooling of compact stars
due to changes in the transport properties and neutrino emissivities
have been investigated much in detail, see
[\cite{Blaschke:1999qx}, \cite{Page:2000wt}, \cite{Blaschke:2000dy},
 \cite{Blaschke:2003yn}], and may
even contribute to the explanation
of the relative low temperature of
the pulsar in the supernova remmant 3C58 [\cite{cooling}].

Unlike the case of normal
(electronic) superconductors, the pairing energy gap in quark matter is
of the order of the Fermi energy so that diquark condensation gives
considerable contributions to the equation of state (EoS) of the order
of $({\Delta}/{\mu})^2$. Therefore, it has been suggested
that there might be scenarios which
identify the unknown source of the energy of $~10^{53}$ erg with a release
of binding energy due to Cooper pairing of quarks in the core of a cooling
protoneutron star [\cite{Hong:2001gt}].
In that work the total diquark condensation energy released in
a bounce of the core is estimated as
$({\Delta}/{\mu})^2M_{{\rm core}}$ corresponding to a few
percent of a solar mass, that is $~10^{52}$ erg.
In this estimate, general relativistic effects have been disregarded. It has
been shown in [\cite{Blaschke:2003yn}]
by solving the selfconsistent problem of the star configurations,
however, that these effects due to the stiffening of the EoS in the diquark
condensation transition lead to an increase in the gravitational mass
of the star  contrary to the naive estimates.

It has also been demonstrated [\cite{Blaschke:2003yn}] that the energy release
due to cooling of a quark core in a protoneutron star does
not occur whithin an explosive process,
since the diquark condensation is a second order phase transition.

In the present work, we propose a new  mechanism of energy release which
involves a first order phase transition induced by antineutrino untrapping.
(Anti-)neutrino trapping occurs in hot compact star configurations at temperatures
$T\geq 1$ MeV where the mean free path of (anti-)neutrinos is smaller
than the typical size of a star [\cite{Prakash:2001rx} and
references therein].

During the collapse in the  hot era of protoneutron star evolution,
antineutrinos are produced due to the $\beta$-processes.
Since they have a  small mean free path, they cannot escape
and the asymmetry in the system is increased.
This entails that the formation of the diquark condensate
is shifted to higher densities or even inhibited depending on the
fraction of trapped antineutrinos.

As the quark star cools, a two-phase structure will occur.
Despite the asymmetry, the interior of the quark star (because of its
large density) could consist of color superconducting quark matter, whereas
in the more dilute outer shell, diquark condensation cannot occur and quark
matter remains in the normal state, opaque to antineutrinos for $T\geq 1$ MeV.
When in the continued cooling process the antineutrino mean free path
increases above the size of this normal
matter shell, an outburst of neutrinos
occurs and gives rise to an energy release of
the order of $10^{51}-10^{52}$ erg.
This untrapping transition is of first order and
could lead to an explosive phenomenon.

The scenario to be detailed in the present paper suggests that  the
first pulse of neutrinos emitted in the deleptonization stage of the
core collapse, after a cooling time scale, shall be followed
by a second pulse of antineutrinos
 as an observable characteristics of the present scenario.

\section{Hot 2SC quark matter}
The investigation of the phase structure of electrically and color neutral 
quark matter in $\beta$-equilibrium within quark models which {\it assume} 
approximate $SU(3)$ flavor symmetry has revealed [\cite{Alford:2002kj}, 
\cite{Steiner:2002gx}] that the Color Flavor Locking (CFL) phase is 
energetically favored over the 2-Flavor Superconductivity (2SC) one.
On the other hand, it has been shown [\cite{Gocke:2001ri}, 
\cite{Neumann:2002jm}] that at low temperatures a sequential deconfinement 
of light and strange quark flavors occurs in {\it dynamical quark models} 
which solve the three-flavor gap equations. 
Thus, strange quark matter phases like the CFL one do appear only 
at high densities in the very
inner core of compact stars and do not occupy a large enough volume in
order to cause observable effects.
Therefore, we consider in the present work two flavor quark matter
in the 2SC phase only.

\subsection{Thermodynamic potential for asymmetric quark matter}

We consider the grand canonical thermodynamic potential
for 2SC quark matter within a nonlocal chiral quark model 
[\cite{Blaschke:2003yn}]
where in the mean field approximation the mass gap $\phi_f$ and the
diquark gap $\Delta$ appear as order parameters and a decomposition
into color ($c\in\{r,b,g\}$) and flavor ($f\in\{u,d\}$) degrees of
freedom can be made.


\begin{eqnarray}
&&
\Omega_q(\{\phi_f\},\Delta;\{\mu_{fc}\},T)=
\sum_{c,f}\Omega^c(\phi_f,\Delta;\mu_{fc},T)~,
\end{eqnarray}
where  $T$ is the  temperature and $\mu_{fc}$
the chemical potential for the quark with flavor $f$ and color $c$.


The contribution of quarks with  given color $c$ and flavor $f$ to the
thermodynamic potential is
\begin{eqnarray}
&&
\Omega^c(\phi_f,\Delta;\mu_{fc},T)+ \Omega_{vac}^c=
\frac{\phi_f^2}{24~G_1}+\frac{\Delta^2}{24~G_2}
\nonumber\\
&&-
\frac{1}{\pi^2}
\int^\infty_0dqq^2
\{
\omega\left[
\epsilon_c(E_f(q)+\mu_{fc}),T
\right]
+
\nonumber\\
&&
\omega\left[
\epsilon_c(E_f(q)-\mu_{fc}),T
\right]
\}~,
\label{ome2}
\end{eqnarray}
where $G_{1}$ and $G_{2}$ are coupling constants in the 
scalar meson and diquark chanels, respectively.
The dispersion relation for unpaired quarks with dynamical mass
function $m_f(q)=m_f+g(q)\phi_f$
is given by
\begin{eqnarray}
&& E_f(q)=
\sqrt{q^2+m^2_f(q)}~.
\end{eqnarray}
In Eq. (\ref{ome2}) we have introduced the notation
\begin{eqnarray}
&&
\omega\left[\epsilon_c,T\right]= T\ln\left
[1+\exp\left(-\frac{\epsilon_c}{T}\right)\right]+\frac{\epsilon_c}{2}~,
\label{ome3}
\end{eqnarray}
where the first argument is  given by
\begin{eqnarray}
&&\epsilon_c(\xi)=\xi\sqrt{1+\Delta^2_c/\xi^2}~.
\end{eqnarray}
When we choose the green and blue colors to be paired
and the red ones to remain unpaired, we have
\begin{eqnarray}
&& \Delta_c=g(q)\Delta(\delta_{c,b}+\delta_{c,g}).
\end{eqnarray}

For a homogeneuos system, the thermodynamic potential
$\Omega_q$ corresponds to the pressure; therefore the constant
$\Omega_{vac}=\sum_{c}\Omega_{vac}^c$ is
chosen such that the pressure of the physical vacuum vanishes.

The nonlocality of the interaction between the quarks
is implemented via formfactor functions
$g(q)$ in the momentum space.
We use the Gaussian formfactor defined as
\begin{eqnarray}
&&
g(q)=\exp(-q^2/\Lambda^2)~,
\end{eqnarray}
since we could show that for this choice stable hybrid stars are possible
[\cite{Blaschke:2003rg}].
The parameters: $\Lambda = 1.025$ GeV, $G_1= 3.761~\Lambda^2$  and
$m_u=m_d=m=2.41$ MeV are fixed by the pion mass $m_{\pi}=140$  MeV,
pion decay
constant $f_{\pi}=93$  MeV and
the constituent quark mass $m_0=330$ MeV
at $T=\mu=0$ [\cite{Schmidt:1994di}]. 
The constant $G_2$ is a free parameter of the approach which we fix as 
$G_2=0.86~ G_1$.


 We introduce new variables:
the quark chemical potential for the color $c$,
$\mu_{qc}=(\mu_{uc}+\mu_{dc})/2$, and the
 chemical potential of the isospin
asymmetry, $\mu_I = (\mu_{uc}-\mu_{dc})/2$, which is color independent.
We consider now symmetric ($\mu_{uc}=\mu_{dc}$) or nearly symmetric
($\mu_{uc} \simeq \mu_{dc}$) quark matter
as the preferable situation to form a
pair of fermions in the color space, i.e. $|\mu_I| < \mu_{qc}$.

The color asymmetry induces a splitting of the quark chemical potentials
relative to the mean value for two flavors $\mu_q$ which is proportional to the
new chemical potential $\mu_8$. Therefore we can write,
\begin{eqnarray}
\mu_{qc}=\mu_q+\frac{\mu_8}{3}(\delta_{c,b}+\delta_{c,g}-2\delta_{c,r})~,
\end{eqnarray}
where $\mu_q$ and $\mu_8$ are conjugate to the quark number
density and the color charge density, respectively.

We perfom the approximation
$\phi_u=\phi_d=\phi$
 and express the thermodynamic potential as in
[\cite{Kiriyama:2001ud}]


\begin{eqnarray} \label{Omeg1}
&&\Omega_q(\phi,\Delta;\mu_{qr},\mu_{qb},\mu_I,T)+\Omega_{vac}
=\frac{\phi^2}{4G_1}+\frac{\Delta^2}{4G_2}
\nonumber\\
& &
-\frac{1}{\pi^2}\int^\infty_0dqq^2\{
\omega\left[\epsilon_r(-\mu_{qr}-\mu_I),T\right]+
\nonumber\\
& &
\omega\left[\epsilon_r(\mu_{qr}-\mu_I),T\right]+
\omega\left[\epsilon_r(-\mu_{qr}+\mu_I),T\right]+
\nonumber\\
& &
\omega\left[\epsilon_r(\mu_{qr}+\mu_I),T\right]
\}\nonumber\\
& &
-\frac{2}{\pi^2}\int^\infty_0dqq^2\{
\omega\left[\epsilon_b(E(q)-\mu_{qb})-\mu_I,T\right]+
\nonumber\\
& &
\omega\left[\epsilon_b(E(q)+\mu_{qb})-\mu_I,T\right]+
\nonumber\\
& &
\omega\left[\epsilon_b(E(q)-\mu_{qb})+\mu_I,T\right]+
\nonumber\\
& &
\omega\left[\epsilon_b(E(q)+\mu_{qb})+\mu_I,T\right]
\}~,
\label{ome9}
\end{eqnarray}

where the factor $2$ in the last integral comes from the degeneracy of the
blue and green colors ($\epsilon_b=\epsilon_g$).

The conditions for the local extrema of $\Omega_q$, correspond to
coupled gap equations for the two order parameters $\phi$ and  $\Delta$
\begin{eqnarray}
&&
{\partial \Omega \over \partial \phi}\bigg|_{\phi=\phi_0,\Delta=\Delta_0}=
{\partial \Omega \over \partial \Delta}\bigg|_{\phi=\phi_0,\Delta=\Delta_0}=0~.
\end{eqnarray}
The global minumum of $\Omega_q$ represents the state of thermodynamical
equilibrium from which all equations of state can be obtained by derivation.

\subsection{Beta equilibrium with electrons and trapped (anti)neutrinos}

The stellar matter in the quark core of compact stars consists of
$u$ and  $d$ quarks, electrons $e$ and antineutrinos  $\bar \nu_e$
under the conditions of

\begin{itemize}
\item
$\beta$-equilibrium:
$d \longleftrightarrow u+e^-+\bar \nu_e$,
which in terms of chemical potentials reads
$$
\mu_e+\mu_{\bar \nu_e}=-2\mu_I~,
$$
\item
charge neutrality: $\frac{2}{3}n_u-\frac{1}{3}n_d-n_e = 0$,
which could also be written as
$$
n_B+n_I-2n_e=0~,
$$
\item
color neutrality:  $n_8=\frac{2}{3}n_{qr}-\frac{1}{3}n_{qb}=0$~.
\end{itemize}
Here the number densities $n_j$ are defined in relation to the corresponding 
chemical potentials $\mu_j$ as
\begin{eqnarray}
&&
n_j=-\frac{\partial \Omega}{\partial \mu_j}\bigg|_{\phi_0,\Delta_0;T}~,
\end{eqnarray}
where the index $j$ denotes the particle species.
The baryon chemical potential defined as $\mu_B = 3\mu_q-\mu_I$.
The solution of the color neutrality condition shows that
$\mu_8$ is about 5-7 MeV in the region of relevant densities
($\mu_q\simeq 300-500$ MeV). Since the isospin asymmetry is independent
of $\mu_8$ we consider $\mu_{qc}\simeq\mu_q$ in our following calculations.

The total thermodynamic potential $\Omega$ contains besides the quark contribution
$\Omega_q$ also that of the leptons $\Omega^{id}$
\begin{eqnarray}
\Omega(\phi,\Delta;\mu_q,\mu_I,\mu_e,\mu_{\bar \nu_e},T)&=&
\Omega_q(\phi,\Delta;\mu_q,\mu_I,T) \nonumber  \\
&+& \sum_{l \in \{e,\bar \nu_e \}}\Omega^{id}(\mu_l,T)~,
\end{eqnarray}
which is assumed to be a massless, ideal Fermi gas
\begin{eqnarray}
&&
\Omega^{id}(\mu,T)=-\frac{1}{12\pi^2}\mu^4-\frac{1}{6}\mu^2T^2-\frac{7}{180}
\pi^2T^4~.
\end{eqnarray}
\begin{figure}[h]
\psfig{figure=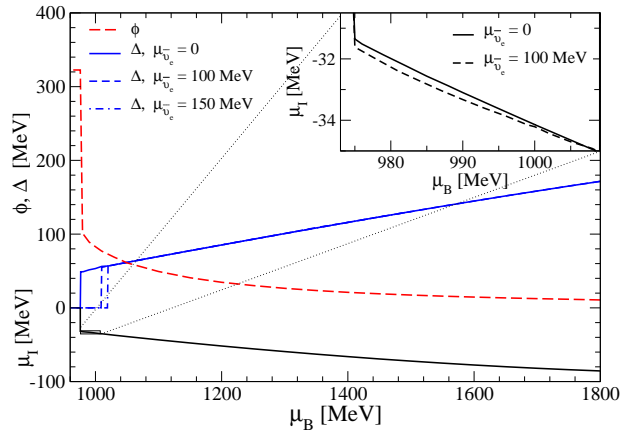,width=0.5 \textwidth,angle=-90}
\caption{Mass gap $\phi$, diquark gap $\Delta$ and isospin chemical
potential $\mu_I$
as a function of the baryon chemical potential $\mu_B$ for different
values of the antineutrino chemical
potential $\mu_{\bar \nu_e}$. Solutions obey $\beta$-equilibrium and
charge neutrality conditions.
\label{GEprue}}
\end{figure}
In Fig. \ref{GEprue} we show the solution of the gap equations for the order
parameters $\phi$  and $\Delta$  as a function
of the baryon chemical potential $\mu_B$ at $T=0$. Also the solution for
the isospin chemical potential $\mu_I$ is plotted.
We can see that with increasing antineutrino chemical potential
$\mu_{\bar \nu_e}$ 
the absolute value of the chemical potential for isospin asymmetry
$|\mu_I|$ increases (see the inset of Fig. \ref{GEprue}). 
We will refer to this value as the asymmetry in the system, 
and when it increases  the onset of the
superconducting phase transition is shifted to higher densities.
For the region of chemical potentials $\mu_q$ where the diquark condensate
is nonvanishing its value coincides with that for the case $\mu_I=0$.
This region, however, shrinks for
$\mu_{\bar\nu_e}^{min}\leq \mu_{\bar\nu_e}$.
The minimum value of antineutrino chemical potential for which the onset of
diquark condensation does not coincide with the chiral transition is 
$\mu_{\bar\nu_e}^{min}=30$ MeV for a density of $\mu_B = 1$ GeV. 

\subsection{Equations of state}
\begin{figure}[h]
\psfig{figure=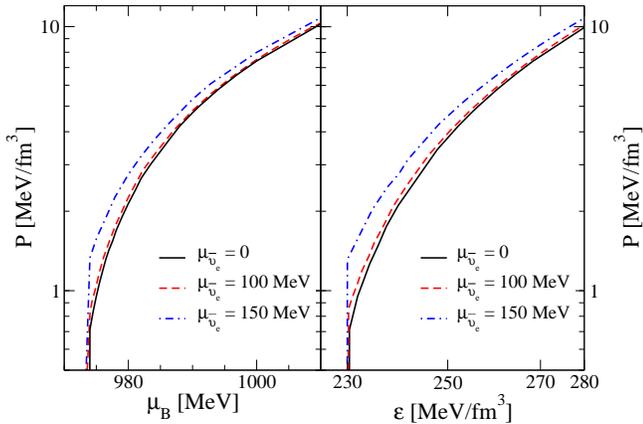,width=0.5 \textwidth,angle=-90}
\caption{Pressure vs. baryon chemical potential (left panel) and 
energy density (right panel)
 for different values of the antineutrino chemical
potential $\mu_{\bar \nu_e}$. Due to antineutrino trapping the onset 
of superconductivity in quark matter is shifted to higher energy 
densities and equation of state becomes {\it harder}.
\label{EoSfig}}
\end{figure}

The equations of state are obtained from the thermodynamic potential
(\ref{ome9}).
For homogeneous systems,
the pressure is $P=-\Omega$,
the entropy density is
$s=-({\partial \Omega}/{\partial T})_{\mu_q}$ and
the energy density is given by the Gibbs fundamental relation
\begin{eqnarray}
&&\varepsilon=sT-P+\mu_qn_q+\mu_In_I.
\end{eqnarray}
In Fig. \ref{EoSfig} we show the pressure as a function of the
baryon chemical potential and of the energy density for different 
values of the antineutrino chemical potential. 
The effect of the presence of trapped antineutrinos is twofold: 
(i) the onset of superconductivity in quark matter
is shifted to higher energy densities
with increasing $\mu_{\bar \nu_e}$ and 
(ii) the equation of state becomes {\it harder}.
%
\section{Hot quark stars and neutrino untrapping}

Hot quark star configurations have been considered first in
[\cite{Kettner:1995zs}] for isothermal configurations and in
[\cite{Blaschke:1998hy}] for adiabatic ones.
When determining the mass defect of a compact star
configuration due to diquark condensation in a cooling process,
it has been shown in [\cite{Blaschke:2003yn}] that general relativistic effects
become important.
Effects which have been estimated by extrapolating
the binding energy per Cooper pair [\cite{Hong:2001gt}] may even be
conterbalanced when the mass distribution is determined selfconsistenly.
This observation underlines the necessity to estimate the release of
binding energy in evolutionary processes of compact stars which change
the equation of state selfconsistently with corresponding changes in the
mass distribution and the gravitational field.
In Ref. [\cite{Blaschke:2003yn}],
however, no trapped (anti)neutrinos have been
considered and it has been demonstrated that in this case the diquark
condensation occurs within a second order phase transition, so that no
sudden release of energy had to be expected.
The situation may change in the case when the transition is inhibited by
the presence of trapped antineutrinos which increase the asymmetry above
a critical value and therefore  prevent diquark condensation.

\subsection{Structure and stability of compact stars including
trapped (anti)neutrinos}

We consider  star configurations which are defined as solutions of the
Tolman-Oppenheimer-Volkoff equations
\begin{eqnarray}
\frac{dP(r)}{dr}&=&
-\frac{[\epsilon(r)+P(r)][m(r)+4\pi r^3P(r)]}{r[r-2m(r)]}~,
\\
m(r)&=&4\pi\int_{0}^{r} r'^2 \epsilon(r') dr'~,
\end{eqnarray}
where $r$ is the distance from the center and $m(r)$ the mass enclosed
in a sphere of radius $r$.

The equations are solved for the set of central quark number
densities $n_{q}$  for which the stars are stable.
The total mass $M = m(R)$ of the star is defined by the radius $R$ being
the distance to the star surface which fulfills the condition of vanishing
pressure $P(R)=0$.

\begin{figure}[bht]
\psfig{figure=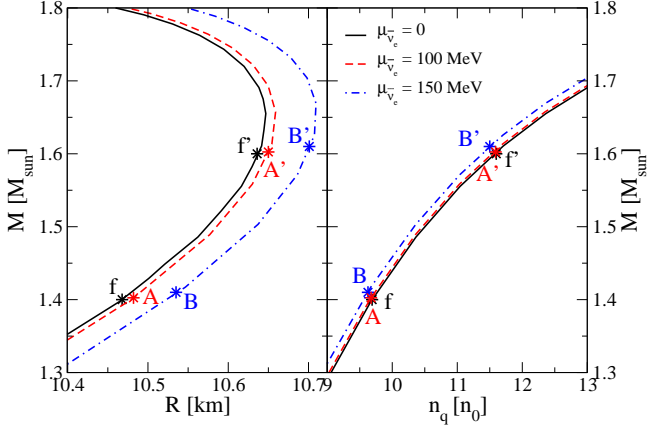,width=0.5 \textwidth,angle=-90}
\caption{Quark star configurations for different antineutrino
chemical potentials $\mu_{\bar \nu_e}=0, ~100,~150$ MeV.
The total mass $M$ in solar masses
($M_{{\rm sun}}\equiv M_\odot$ in the text)
is shown as a function of the radius $R$ (left panel) and
of the central number density  $n_q$ in units of the
nuclear saturation density $n_0$ (right panel).
Asterisks denote two different sets of configurations (A,B,f) and (A',B',f')
with a fixed total baryon number of the set.
\label{QSConf}}
\end{figure}
The total mass as a function of radius and central quark number
density of the configurations is plotted in Fig. \ref{QSConf}
for the case without neutrinos $\mu_{\bar \nu_e}=0$ (solid line) and for
two cases with finite antineutrino chemical potential:
$\mu_{\bar \nu_e}=100$ MeV (dashed line) and 
$\mu_{\bar \nu_e}=150$  MeV (dash-dotted line).
The latter value relates to our estimate for the maximum antineutrino number 
density which should occur just at the deconfinement phase transition, but 
still larger values could also be possible [\cite{Steiner:2002gx}].
From Fig. \ref{EoSfig} we can see that the EoS without 
antineutrinos is softer than that with antineutrinos (it has a lower pressure 
at a given energy density) and in Fig.\ref{QSConf} we show that this
 allows more compact configurations (left panel).
Consequentely, the presence of antineutrinos tends to increase  
the mass of the star for
a given central density, see the right panel of Fig.\ref{QSConf}.

In order to estimate the effect of antineutrinos on the mass of 
star configurations we choose a reference configuration  without
antineutrinos with the mass of a typical neutron star $M_f = 1.4~ M_{\odot}$,
see Fig. \ref{QSConf}. The corresponding radius is
$R_f = 10.47$ km and the central density $n_q = 9.69~n_0$,
where $n_0 = 0.16~{\rm fm}^{-3}$
is the saturation density of nuclear matter.
The configurations with trapped antineutrinos and nonvanishing
$\mu_{\bar\nu_e}$ to compare with are chosen to have the same total baryon
number as the reference star: $N_B = 1.51~ N_{\odot}$, where
$N_{\odot}$ is the total baryon number of the sun.
For $\mu_{\bar\nu_e}=100$ MeV we obtain $M_A = 1.4025~M_{\odot}$ whereas 
for $\mu_{\bar\nu_e}=150$ MeV we have $M_B = 1.41~M_{\odot}$.
The differences in the radii are  $R_A - R_f= 0.01$ km and $R_B-R_f = 0.06$ km
and the changes in the central densities $n_q^A-n_q^f = -0.016~n_0$ and
$n_q^B-n_q^f = -0.064~n_0$, respectively. This is a consequence of the
hardening of the EoS due to the presence of antineutrinos.
A second set of configurations ($A',B',f'$) with a fixed baryon number 
and $M_f=1.6~M_{\odot}$ is shown in the  Fig.\ref{QSConf}. 

The mass defect $\Delta M_{if} = M_i - M_f$ can be interpreted as an energy
release if there is a process which relates the configurations with $M_i$ and
$M_f$ being the initial and final states, respectively.
In the following Subsection we discuss a possible evolution
scenario and we calculate the corresponding energy release.

\subsection{Estimate of energy release in antineutrino untrapping}

The antineutrino chemical potential $\mu_{\bar \nu_e}$ increases the asymmetry
of the system and can prevent the diquark condensation provided the critical
value $\mu_{\bar \nu_e}$  is exceeded, see Fig.\ref{GEprue}.
The inhibition of diquark condensation could provide conditions for the
explosive release of an amount of energy coming from the mass defect
introduced in the previous subsection.

During the collapse  of a protoneutron star, neutronization takes place
via the inverse $\beta$-process releasing neutrinos which can escape
through the yet dilute and cold outer shell of the protoneutron star.
In the continuation of the collapse the temperature and the density
increase in the protoneutron star core where the proton fraction rises again.
Antineutrinos are created by the direct $\beta$-process in hot and dense matter
and cannot escape since their mean free path is
much smaller than the core radius.
When under these conditions the deconfinement phase transition to quark matter
occurs, this has two important consequences: (i) due to high temperatures of
the order of $40$ MeV the critical density of the phase transition is much
lower than at zero temperature and thus a larger fraction of the star will be
in the new phase and (ii) due to {\it antineutrino trapping} the cooling is
delayed.

The star cools down by surface emission of photons and antineutrinos.
The region of the star where the temperature falls below the density dependent
critical value for diquark condensation, will transform to the color
superconducting state which is almost transparent to (anti)neutrinos.
But nevertheless due to the trapped antineutrinos there is a dilute normal
quark matter shell which prevents neutrino escape from the
superconducting bulk of the star.
The criterion for the  neutrino untrapping transition is to cool the star
below a temperature of about 1 MeV when the mean free path of neutrinos
becomes larger than the shell radius [\cite{Prakash:2001rx}].
If at this temperature the antineutrino chemical potential is still large
then the neutrinos can escape in a sudden outburst. If it is small then there
will be only a gradual increase in the luminosity.
An estimate for the possible release of energy within the outburst scenario
can be given via the mass defect  defined in the previous subsection
between an initial configuration with trapped neutrinos (state $A$ or $B$) and
a final configuration without neutrinos (state $f$).
\begin{figure}[h]
\psfig{figure=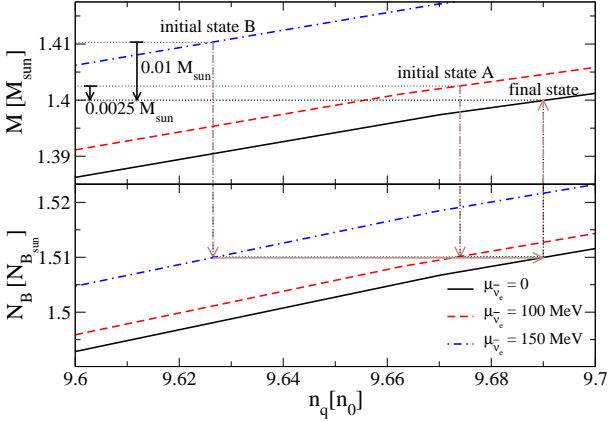,width=0.5 \textwidth,angle=-90}
\caption{Quark star configurations with diquark condensation
 as a function of
the central number density $n_q$ in units of the
nuclear number density $n_0$.
The mass defect for the transition
from initial configurations with $\mu_{\bar \nu_e}=100,~150$ MeV
to a final configuration with $\mu_{\bar \nu_e}=0$
at constant total baryon number is shown.
\label{QSevol}}
\end{figure}
In Fig. \ref{QSevol} the integral characteristics of the star
configuration: the total mass $M$ and the total baryon number $N_B$
are plotted as a function of the central number density $n_q$.
We take two possible initial states ($A$, $B$ with $M_A$, $M_B$ respectively)
with different antineutrino chemical potentials and we make the
evolution to a final state (without antineutrinos, with $M_f$)
at constant total baryon number $N_B$ as we show in Fig. \ref{QSConf}.
The configurations with trapped antineutrinos 
($\mu_{\bar\nu_e}= 100,~150 $ MeV)
correspond to possible initial states before the outburst and the 
configuration without antineutrinos ($\mu_{\bar \nu_e}=0$) is the final 
state after it.
The mass defect ranges from
$\Delta M_{Af} = 0.25~\%~M_{\odot}$
to $\Delta M_{Bf} = 1~\%~M_{\odot}$ for the initial states  $A$ and $B$,
respectively.
We see that the release of energy can be as big as about
$1.5~ \%$ of the initial mass.
\begin{figure}[h]
\psfig{figure=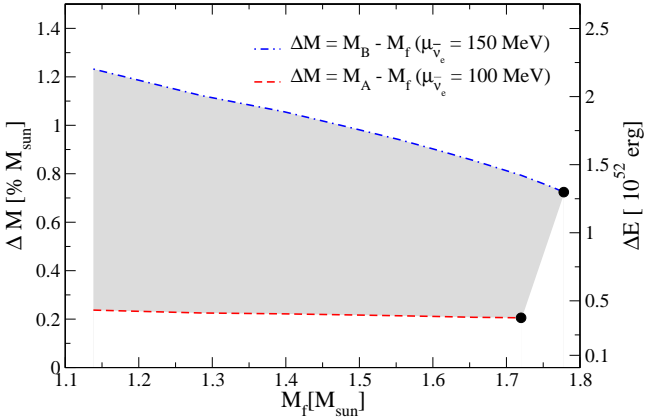,width=0.5 \textwidth,angle=-90}
\caption{Mass defect $\Delta M$ and corresponding energy release $\Delta E$
due to antineutrino untrapping
as a function of the mass of the final state $M_f$.
The shaded region is defined by the estimates for the upper and lower
limits of the antineutrino chemical potential
in the initial state $\mu_{\bar \nu_e}=150$ MeV (dashed-dotted line) and
$\mu_{\bar \nu_e}=100$ MeV (dashed line), respectively.
\label{Deltamass}}
\end{figure}

We can make the same construction for different final states and
in Fig. \ref{Deltamass} we show the mass defect and the corresponding
energy release as a function of the mass of the final configuration.
Two lines for different initial states
are plotted: the dashed one for $A$ configurations
(with $\mu_{\bar \nu_e}=100$ MeV) and the dash-dotted one for
$B$ configurations (with $\mu_{\bar \nu_e}=150$ MeV). The final state
$f$ with vanishing $\mu_{\bar \nu_e}$ has the mass $M_f$ which defines
the conserved baryon number involved in the untrapping transition.
The dots represent  the end points of the stable configurations.
The shaded region in between both curves
shows the possible mass defect (or corresponding energy release)
according to our estimate of the range of the
antineutrino chemical potentials considered.

Since the estimated energy release is as large as $10^{51}-10^{52}$ erg
the scenario suggested in the present paper could be discussed as a possible
engine driving supernova explosions and gamma ray bursts.

A second pulse of antineutrinos is expected to be a signal of the
untrapping transition.

\section{Conclusions}

We have investigated the effects of
trapped antineutrinos on the asymmetry
and  diquark condensates in a quark star configurations.
By comparing configurations with fixed baryon number the
release of energy in an antineutrino untrapping transition
is estimated to be of the order of $10^{52}$ erg.
Such a transition is of first order so that antineutrinos can be
released in a sudden process (burst). This scenario could play an important
r\^ole to solve the problem of the engine of
supernova explosions and gamma ray bursts.
A second antineutrino pulse is suggested as an observable characteristics of
the present scenario.

\begin{acknowledgements}
Research of D.N.A. was supported in part by the CONICET PIP 03072 (Argentina),
by the DFG GK 567 ``Stark korrelierte Vielteilchensysteme'' (Rostock University), by DAAD grant No. A/01/17862 and by Landesgraduiertenf\"orderung von Mecklenburg Vorpommern (Germany).
H.G. acknowledges support by DFG under grant No. 436 ARM 17/5/01.
D.N.A. and H.G. acknowledge the hospitality of the Department of Physics at
the University of Rostock where this research has been performed.
\end{acknowledgements}

\bibliographystyle{aa} 
\bibliography{abg2} 
\end{document}